\documentstyle[12pt,psfig]{article}
\def\({\left(}
\def\){\right)}
\def\lb{\left|}
\def\rb{\right|}
\def\<{\left<}
\def\>{\right>}
\def\[{\left[}
\def\]{\right]}
\def\ll{\left\{}
\def\lr{\right\}}
	
\def\journal#1#2#3#4{{\it #1} {\bf #2} (#3) #4}
\def\prl{Phys. Rev. Lett.}
\def\pl{Phys. Lett.}
\def\np{Nucl. Phys.}

\def\pr{Phys. Rev.}

\def\r2{\sqrt{2}}
\def\vhlp{V_{h\ell^\prime}}

\def\lg{{\cal L}}
\def\acp{{\cal A}_{\rm CP}}
\def\br{{\cal BR}}
\def\b{{\cal B}}
\def\sq{{\tilde{q}}}
\def\sd{{\tilde{d}}}
\def\su{{\tilde{u}}}
\def\sl{{\tilde{\ell}}}
\def\sn{{\tilde{\nu}}}
\def\sf{{\tilde{f}}}
\def\im{{\rm Im}}
\def\re{{\rm Re}}
\def\m{{\cal M}}
\def\t{{\cal T}}
\def\f{{\cal F}}
\def\hlx{q_h \rightarrow q_\ell \, X_{jj}}
\def\tuxd{t \rightarrow u(c) \, X_{d_j\bar{d}_j}}

\def\bdxu{b \rightarrow d(s) \, X_{u_j\bar{u}_j}}

\def\tuxi{t \rightarrow u(c) \, X_{jj}}
\def\bdxi{b \rightarrow d(s) \, X_{jj}}
\def\cuxi{c \rightarrow u \, X_{jj}}
\def\sdxi{s \rightarrow d \, X_{jj}}
\def\hla{q_h \rightarrow q_{\ell^\prime} \, X^\pm}
\def\tbw{t \rightarrow b \, W}

\def\qh{q_h}
\def\l{\ell}
\def\lep{{\ell^\prime}}
\def\hp{{h^\prime}}
\def\ql{q_\ell}
\def\qb{\bar{q}}
\def\w{{\cal W}}
\def\rpv{{R_{\rm p}\hspace{-9.5pt}/}}
\def\rp{R_{\rm p}}
\def\lijk{\lambda_{ijk}}

\def\lp{{\lambda^\prime}}
\def\lpijk{{\lambda_{ijk}^\prime}}

\def\lpp{{\lambda^{\prime \prime}}}
\def\lppijk{{\lambda_{ijk}^{\prime \prime}}}

\def\fx{{f_{X_{jj}}}}
\def\fxh{{\hat{f}_{X_{jj}}}}

\def\ex{{\epsilon_{X_{jj}}}}
\def\xu{{X_{u_j\bar{u}_j}}}
\def\xd{{X_{d_j\bar{d}_j}}}
\def\xi{{X_{jj}}}
\def\mh{{{m}_h}}

\def\mx{{\hat{m}_x}}
\def\my{{\hat{m}_y}}
\def\mz{{\hat{m}_z}}

\def\mw{{m_W}}
\def\mwh{{\hat{m}_W}}

\def\gxy{{g^{xy}}}
\def\gfh{{\hat{G}_F}}
\def\cli{{\cal C}_{\ell j\b}}
\def\chi{{\cal C}_{hj\b}}
\def\clif{{\cal C}_{\ell j\sf}}
\def\chif{{\cal C}_{hj\sf}}
\def\clifp{{\cal C}_{\ell j\sf^\prime}}
\def\chifp{{\cal C}_{hj\sf^\prime}}
\def\clibx{{\cal C}_{\ell j\b_x}}
\def\chibx{{\cal C}_{hj\b_x}}
\def\cliby{{\cal C}_{\ell j\b_y}}
\def\chiby{{\cal C}_{hj\b_y}}
\def\clibh{{\cal C}_{\ell \hp\b}}
\def\chibh{{\cal C}_{h\hp \b}}

\title{Direct CP violation in semi-inclusive flavor-changing
	neutral current decays in the MSSM without $R-$parity}
\author{\vspace{5mm}\\
        {\bf L. T. Handoko}$^{1,2}$\thanks{
	E-mail address : lthandoko@bigfoot.com} 
	\hspace{2mm} and \hspace{2mm} 
	{\bf J. Hashida}$^2$\thanks{
	E-mail address : jhashida@kakuri4-pc.phys.sci.hiroshima-u.ac.jp}\\
	\vspace{2mm}\\
	$^1$Laboratory for Theoretical Physics and Mathematics \\
	Indonesian Institute of Sciences \\
	Kom. Puspitek Serpong P3FT--LIPI, Tangerang 15310, Indonesia\\
	\vspace{1mm}\\
	$^2$Department of Physics, Hiroshima University \\
        1-3-1 Kagamiyama, Higashi Hiroshima 739-0046, Japan}
\date{}

\begin{document}

\maketitle
\begin{picture}(0,0)
       \put(310,340){LFTMLIPI-139802}
       \put(310,325){HUPD-9802}
       \put(310,310){February 1998}
\end{picture}

\thispagestyle{empty}

\begin{abstract}
Semi-inclusive decays, $\hlx$, are studied in the framework 
of the minimal supersymmetric standard model without $R-$parity, 
where $\qh$ ($\ql$) are the second or third (first or second) 
generation quarks with the same charge and $\xi$ is a vector 
meson formed by $q_j \qb_j$. The study is focused on the
contributions of sfermions with $m_{\sf} < m_{\rm top}$. 
In this mass region, CP asymmetries in top decays can 
be induced by taking into account the decay-widths of the 
exchanged-bosons, while in light-quark decays it can be generated 
due to the long-distance effects. The contributions of sfermions 
also alter the branching-ratios destructively or constructively 
depend on the phases of complex couplings of the $R-$parity 
violation interactions.
\end{abstract}

\vspace*{5mm}
PACS number(s) : 14.65.Ha, 14.65.Fy

\clearpage

The discovery of top quark \cite{top} have accomplished the 
particle content of quark sector predicted in the standard model 
(SM). The heavy top and also other light-quarks have been 
attracted by theorists and experimentalists to test 
the SM as well as open a window for the 
physics beyond the SM. Some valuable information are expected 
from some classes of its decays that should be observed in, 
namely, top and meson factories. Theoretically, the $BDK$ meson 
decays have been studied briskly from a few decades ago, while 
most of top decays are still under studied. 

Comparing both of them, the study of $BDK$ decays are 
mostly confronted with theoretical difficulties like the 
non-perturbative effects. On the other hand, in top decays 
the difficulties are almost coming from the 
experimental side that is still far to carry out some 
precise measurements as will be achieved for $BDK$ decays in 
the meson factories, although top decays are clean of 
theoretical uncertainties because of its large mass scale.
It is also well known that most of models beyond the SM 
contribute significantly to the rare $BDK$ decays, and 
its presence should be examined in the present or  
near future meson factories. On the contrary, although 
the rare top decays are also very sensitive to the new 
physics, e.g. \cite{2hdm,mssm,mssmrp}, the rates are still at 
unreachable level even in the future top factories as 
the upgraded Tevatron or LHC. These facts encourage 
us to consider some modes with the order between the 
lowest charged-current and the rare decays. Then we consider
the class of semi-inclusive decays $\hlx$. Here $\qh$ 
($\ql$) are the second or third (first or second) 
generation quarks and have the same charge ($Q_h = Q_\ell$), 
while $\xi$ is any vector meson formed by $q_j \qb_j$. 
Since diagramatically both top and light-quark processes are 
same and the interactions work on them may be related each other, we 
are going to consider both top and light-quark decays 
simultaneously. Definitely, we will discuss the 
flavor-changing semi-inclusive decays : $\tuxi$, $\cuxi$, 
$\bdxi$ and $\sdxi$. The vector mesons are, for example, 
$\xu = \rho, \omega, J/\psi, \cdots$ and 
$\xd = \phi, \Upsilon, \cdots$.

Moreover, in the present paper, the interest is focused 
on the direct CP asymmetry defined as
\begin{equation}
        \acp \equiv \frac{
		\Gamma - \bar{\Gamma}}{
		\Gamma + \bar{\Gamma}} 
                \equiv \frac{\Delta}{\Sigma} \; ,
        \label{eqn:a}
\end{equation}
with $\bar{\Gamma}$ is the complex conjugate of decay-width 
$\Gamma$, while in general 
\begin{eqnarray}
        \Delta & = & -\sum_{x\not=y} 
        	\im \( {\alpha_x}^\ast \, \alpha_y \) \,
                \im \( {\m_x}^\ast \, \m_y \) \; ,
        \label{eqn:delta}\\
        \Sigma & = & 
	\sum_x \lb \alpha_x \rb^2 \lb \m_x \rb^2 + 
	\sum_{x\not=y} \re \( \alpha_x^\ast \, \alpha_y \) \, 
	\re \( \m_x^\ast \, \m_y \) \; ,
        \label{eqn:sigma}
\end{eqnarray}
if one describes the amplitude as $\m \equiv \sum_x \alpha_x \, \m_x$.
Hence, the imaginary parts of $\( {\alpha_x}^\ast \, \alpha_y \)$ 
and $\( {\m_x}^\ast \, \m_y \)$ are required to be non-zero 
coincidently in order to have non-zero CP asymmetry. 

Indeed, in the framework of SM, the CP violation in 
these decays have been studied in some papers. The decays 
$\bdxu$ have been discussed in \cite{bdx} for the typical 
one $b \rightarrow d \, J/\psi$. It has been concluded that 
the CP asymmetry is tiny, i.e. $\sim O({10}^{-3})$, generated due 
to strong or electromagnetic scattering in the final state.
However, the size could be at a few percents level if one takes 
into account the long-distance effects of the intermediate 
states with same quark contents as the final state \cite{fsi}. 
On the other hand, recently the decays $\tuxd$ have also 
been examined in \cite{tux}. Different with the bottom one, 
in the case of top decays, CP violation is induced only by the 
scattering in the final state. It gives the size to be less than 
$O({10}^{-2})$. Therefore in the SM, CP asymmetries in the 
present class of decays are almost at unreachable level of 
experiments, but inversely it makes them to be good probes 
to detect new contributions beyond the SM. 

Presently, one of the well-known candidates for models beyond
the SM is the supersymmetric standard model (MSSM). The model
is attractive because of solution of the naturalness problem
and also a lot of interesting properties. Especially, one of
them that is relevant with our interest is CP violation due 
to the broken $R-$parity ($\rp$). The $\rp-$conservation is 
imposed to prevent the terms which explicitly break the baryon 
($B$) and lepton ($L$) numbers.
In the SM, the gauge symmetry leads to the conservation of 
$B$ and $L$, while in the SUSY model it does not prevent the
terms \cite{susywr}. Without $\rp$, there will be additional
terms in superpotential \cite{susywrfcnc}, that is
\begin{equation}
	\w_\rpv = \lijk \, L_i \, L_j \, E_k^c 
	+ \lpijk \, L_i \, Q_j \, D_k^c 
	+ \lppijk \, U_i^c \, D_j^c \, D_k^c 
	+ \kappa_i \, L_i \, H_2 \; .
	\label{eqn:spot}
\end{equation}
Here, $L$ and $E^c$ ($Q$ and $U^c$, $D^c$) are the lepton doublet 
and anti-lepton singlet (quark doublet and anti-quark singlet) 
left-chiral superfields, while $H_{1,2}$ are the Higgs doublet 
chiral superfields. ($i, j , k$) are generation indices,
while ($\lambda, \lp, \lpp$) are Yukawa coupling-strengths.
In fact, up to now there is still no theoretical preference 
between conserved and violated $\rp$. Phenomenologically, 
some authors have payed attention to these terms, 
especially the stop contribution in above $\lp$ term, because 
of its possibility to explain the HERA anomaly \cite{hera}. 
If this is a true story, then one is required to confirm
the $\rp-$violation in other modes. In this meaning, it is 
worthwhile to adopt the present mode as a probe to test it. 

It is obvious that the semi-inclusive decays, $\hlx$, can 
be induced by either $\lp$ or $\lpp$ terms at tree-level.
Then, the lagrangians that are relevant with the 
present processes are given by expanding the terms in 
Eq. (\ref{eqn:spot}),
\begin{eqnarray}
	\lg_\lp & = & -\lpijk \, \[ 
		{\tilde{\nu}_L}^i \, {\bar{d}_R}^k \, {d_L}^j + 
		{\tilde{d}_L}^k \, {\bar{d}_R}^k \, {{\nu}_L}^i + 
		\({\tilde{d}_R}^j\)^\ast \, \({\bar{\nu}_L}^i\)^c \, 
			{d_L}^j 
		\right. 
		\nonumber \\
	& & \left. \; \; \; \; \; \; \; \; \; \; \; \; - 
		{\tilde{e}_L}^i \, {\bar{d}_R}^k \, {u_L}^j - 
		{\tilde{u}_L}^j \, {\bar{d}_R}^k \, {e_L}^i - 
		\({\tilde{d}_R}^k\)^\ast \, \({\bar{e}_L}^i\)^c \, 
			{u_L}^j 
		\] + {\rm h.c.}  \; ,
	\label{eqn:lglp}\\
	\lg_\lpp  & = & -\lppijk \, \[ 
		{\tilde{d}_R}^k \, \( {\bar{u}_L}^i \)^c \, {d_L}^j + 
		{\tilde{d}_R}^j \, \( {\bar{d}_L}^k \)^c \, {u_L}^i + 
		{\tilde{u}_R}^i \, \( {\bar{d}_L}^j \)^c \, {d_L}^k 
		\] + {\rm h.c.}  \; .
	\label{eqn:lglpp}
\end{eqnarray}
Note that $\lpp$ is anti-symmetric under the interchanges 
of $[j,k]$. These terms induce new contributions to 
the decays with same level as the standard $W-$boson mediated 
diagram. In point of view of the SM, there are two types of 
the decays that may occur in the present model, 
\begin{enumerate}
	\item The SM favored modes induced by $W-$boson, 
		sleptons and down-type squarks exchange diagrams.
	\item The SM forbidden modes induced by sneutrinos 
		and up-type squarks exchange diagrams.
\end{enumerate}
The second one above is known as the tree-level flavor-changing 
neutral current (FCNC) modes that are also allowed in some models 
with additional isosinglet charge (-1/3) quarks. Most important 
difference is, in the present model the unitarity of CKM matrix 
is not altered at all. In this meaning, the mode is very 
interesting, if experimentally the unitarity of CKM matrix is 
known to be conserved while, for example, the decay 
$b \rightarrow s \, \phi$ is observed at appropriate level.
Again, there is no tree-level FCNC modes in up-type quark 
sector in the present model.

After performing Fierz transformation, the amplitude in the 
processes through multiple $\b-$boson mediated diagrams, are 
governed by the following operator 
\begin{equation}
        \m = 
	\r2 \, G_F \, \fx \, m_\xi \, \mw^2 \, {\ex^\mu}^\ast \,
	\sum_{\b} \[ 
	\( \cli^\ast \, \chi \) \, {\f_2}^{\l \xi \b} \] \, 
        \[ \qb_\l \, \gamma_\mu \, L \, q_h \] \; ,
        \label{eqn:m0}
\end{equation}
by taking the factor of SM-like contribution as normalization 
factor. Here $L = {(1 - \gamma_5)}/2$ and ${\f_2}^{\l \xi \b}$ 
is $\b-$boson propagator that will be given later. The vector 
meson is factorized as 
\begin{equation}
        \< 0 \lb \qb_j \, \gamma^\mu \, q_j \rb \xi \> 
                = m_\xi \, \fx \, \ex^\mu \; ,
        \label{eqn:fv}
\end{equation}
where $\fx$ is a constant with dimensions of mass and $\ex^\mu$ 
is the polarization vector. The coupling constants 
$\cli^\ast \, \chi$ are given in Tab. \ref{tab:cc}, where 
$a \equiv \( 2 \, \r2 \, G_F \, \mw^2 \)^{-1}$ and 
$V$ denotes the CKM matrix respectively. In the table, 
non-zero conditions are derived from the anti-symmetric 
of interchanging the indices of $\lpp$, while the allowed 
modes are determined from the kinematics.

\begin{table}[t]
	\begin{center}
	\begin{tabular}{clccccc}
	\hline \hline
	Type & \multicolumn{1}{c}{Decay mode} & $\b$ 
		& $\cli^\ast \, \chi$ & 
		\begin{tabular}{c}
			Non-zero \\
			condition
		\end{tabular} & 
		& 
		\begin{tabular}{c}
			Allowed \\
			mode
		\end{tabular} \\
	\hline \hline 
	1 & $t \rightarrow u \, \xd$ & 
		$\ll \begin{array}{c} 
			W \\
			\sd_k \\
			\sl_k
		\end{array} \right.$ & 
		\begin{tabular}{c}
			${V_{uj}}^\ast \, V_{tj}$ \\
			$a \( {\lpp_{1kj}}^\ast \, \lpp_{3kj} \)$ \\
			$a \( {\lp_{k1j}}^\ast \, \lp_{k3j} \)$
		\end{tabular} & 
		\begin{tabular}{c}
			-- \\
			$k \not= j$ \\
			--
		\end{tabular} & 
		$\left. \begin{array}{c} 
			\\
			\\
			
		\end{array} \lr$ & 
		$j = 1,2,3$ \\
	& $t \rightarrow c \, \xd$ & 
		$\ll \begin{array}{c} 
			W \\
			\sd_k \\
			\sl_k
		\end{array} \right.$ & 
		\begin{tabular}{c}
			${V_{cj}}^\ast \, V_{tj}$ \\
			$a \( {\lpp_{2kj}}^\ast \, \lpp_{3kj} \)$ \\
			$a \( {\lp_{k2j}}^\ast \, \lp_{k3j} \)$
		\end{tabular} & 
		\begin{tabular}{c}
			-- \\
			$k \not= j$ \\
			--
		\end{tabular} & 
		$\left. \begin{array}{c} 
			\\
			\\
			
		\end{array} \lr$ & 
		$j = 1,2,3$ \\
	& $c \rightarrow u \, \xd$ & 
		$\ll \begin{array}{c} 
			W \\
			\sd_k \\
			\sl_k
		\end{array} \right.$ & 
		\begin{tabular}{c}
			${V_{uj}}^\ast \, V_{tj}$ \\
			$a \( {\lpp_{1kj}}^\ast \, \lpp_{2kj} \)$ \\
			$a \( {\lp_{k1j}}^\ast \, \lp_{k2j} \)$
		\end{tabular} & 
		\begin{tabular}{c}
			-- \\
			$k \not= j$ \\
			--
		\end{tabular} & 
		$\left. \begin{array}{c} 
			\\
			\\
			
		\end{array} \lr$ & 
		$j = 1,2$ \\
	& $b \rightarrow d \, \xu$ & 
		$\ll \begin{array}{c} 
			W \\
			\sd_k \\
			\sl_k
		\end{array} \right.$ & 
		\begin{tabular}{c}
			${V_{jd}}^\ast \, V_{jb}$ \\
			$a \( {\lpp_{j1k}}^\ast \, \lpp_{j3k} \)$ \\
			$a \( {\lp_{kj3}}^\ast \, \lp_{kj1} \)$
		\end{tabular} & 
		\begin{tabular}{c}
			-- \\
			$k = 2$ \\
			--
		\end{tabular} & 
		$\left. \begin{array}{c} 
			\\
			\\
			
		\end{array} \lr$ & 
		$j = 1,2$ \\
	& $b \rightarrow s \, \xu$ & 
		$\ll \begin{array}{c} 
			W \\
			\sd_k \\
			\sl_k
		\end{array} \right.$ & 
		\begin{tabular}{c}
			${V_{js}}^\ast \, V_{jb}$ \\
			$a \( {\lpp_{j2k}}^\ast \, \lpp_{j3k} \)$ \\
			$a \( {\lp_{kj3}}^\ast \, \lp_{kj2} \)$
		\end{tabular} & 
		\begin{tabular}{c}
			-- \\
			$k = 1$ \\
			--
		\end{tabular} & 
		$\left. \begin{array}{c} 
			\\
			\\
			
		\end{array} \lr$ & 
		$j = 1,2$ \\
	& $s \rightarrow d \, \xu$ & 
		$\ll \begin{array}{c} 
			W \\
			\sd_k \\
			\sl_k
		\end{array} \right.$ & 
		\begin{tabular}{c}
			${V_{jd}}^\ast \, V_{js}$ \\
			$a \( {\lpp_{j1k}}^\ast \, \lpp_{j2k} \)$ \\
			$a \( {\lp_{kj2}}^\ast \, \lp_{kj1} \)$
		\end{tabular} & 
		\begin{tabular}{c}
			-- \\
			$k = 3$ \\
			--
		\end{tabular} & 
		$\left. \begin{array}{c} 
			\\
			\\
			
		\end{array} \lr$ & 
		$j = 1$ \\
	\hline 
	2 & $b \rightarrow d \, \xd$ & 
		$\ll \begin{array}{c} 
			\su_k \\
			\sn_k
		\end{array} \right.$ & 
		\begin{tabular}{c}
			$a \( {\lpp_{j1k}}^\ast \, \lpp_{j3k} \)$ \\
			$a \( {\lp_{k1j}}^\ast \, \lp_{k3j} \)$
		\end{tabular} & 
		\begin{tabular}{c}
			$k = 2$  \\ 
			--
		\end{tabular} & 
		$\left. \begin{array}{c} 
			\\
			
		\end{array} \lr$ & 
		$j = 1,2$ \\
	& $b \rightarrow s \, \xd$ & 
		$\ll \begin{array}{c} 
			\su_k \\
			\sn_k
		\end{array} \right.$ & 
		\begin{tabular}{c}
			$a \( {\lpp_{j2k}}^\ast \, \lpp_{j3k} \)$ \\
			$a \( {\lp_{k2j}}^\ast \, \lp_{k3j} \)$
		\end{tabular} & 
		\begin{tabular}{c}
			$k = 1$  \\
			--
		\end{tabular} & 
		$\left. \begin{array}{c} 
			\\
			
		\end{array} \lr$ & 
		$j = 1,2$ \\
	& $s \rightarrow d \, \xd$ & 
		$\ll \begin{array}{c} 
			\su_k \\
			\sn_k
		\end{array} \right.$ & 
		\begin{tabular}{c}
			$a \( {\lpp_{j1k}}^\ast \, \lpp_{j2k} \)$ \\
			$a \( {\lp_{k1j}}^\ast \, \lp_{k2j} \)$
		\end{tabular} & 
		\begin{tabular}{c}
			$k = 3$  \\
			--
		\end{tabular} & 
		$\left. \begin{array}{c} 
			\\
			
		\end{array} \lr$ & 
		$j = 1$ \\
	\hline
	\end{tabular}
	\caption{The coupling-strength $\cli^\ast \, \chi$ for 
		each mode.}
	\label{tab:cc}
	\end{center}
\end{table}

First, let us consider the branching-ratio. In general, 
it is better to consider the charged-current decay normalized 
one, that is
\begin{equation}
        \br(\hlx) = \frac{\Gamma(\hlx)}{\Gamma(\hla)}
		\times \br(\hla) \; ,
        \label{eqn:br}
\end{equation}
to eliminate some uncertainties in the overwhole factors.
Here $q_{\ell^\prime}$ is any light-quark that has different 
charge with $q_h$, while $X^\pm$ is anything with charge $\pm1$. 
Definitely, we normalize $\tuxi$ (other light-quark modes) 
with $\tbw$ (its semi-leptonic decays) as usual. The decay-width 
in numerator can be written as 
\begin{equation}
        \Gamma (\hlx) = 
		\frac{\gfh^2 \, \fxh^2 \, \mwh^4 \, \mh}{8 \, \pi} \, 
		\sqrt{g^{\l \xi}} \, {\f_1}^{\l \xi} \, 
		\lb \sum_\b \( \cli^\ast \, \chi \) \, 
		{\f_2}^{\l \xi \b} \rb^2 \; ,
        \label{eqn:dwhlx}
\end{equation}
in the heavy quark center-mass system under assumptions that
$E_\xi = 2 \, E_j$ and $m_\xi = 2 \, m_j$, with $E$ denotes 
time-component of four-momentum. Here, a caret means 
normalization with $\mh$. Keeping the light-quark masses 
\begin{eqnarray}
	\gxy & \equiv & 1 + \mx^4 + \my^4 - 
		2 \, \( \mx^2 + \my^2 + \mx^2 \, \my^2 \) \; ,
        \label{eqn:gxy} \\
        {\f_1}^{xy} & \equiv & 
         \( 1 - \mx^2 \)^2 + \( 1 + \mx^2 \) \, \my^2 - 2 \, \my^4 \; .
        \label{eqn:f1}
\end{eqnarray}
${\f_2}^{\l \xi \b}$ is the $\b-$boson propagator contribution,  
\begin{equation}
        {\f_2}^{xyz} \ll 
	\begin{array}{lcl}
	= \[ \( 1 + \frac{1}{4} \my^2 - 
        \frac{1}{2} \sqrt{\gxy + 4 \, \my^2} - \mz^2 \) 
                + i \, \hat{\Gamma}_z \, \mz \]^{-1} & 
		 {\rm for} & q_h = t \; , \\
	\approx - \mz^{-2} & {\rm for} & q_h \not= t \; , 
	\end{array}
	\right.
        \label{eqn:f2}
\end{equation}
under the kinematical condition : $m_\b < m_h$ for $q_h = t$.
Beware of including the decay-width in the propagator is 
essential for CP asymmetry. This point will be discussed later. 
Further, the charged-current decays in denominator are given
as follows, 
\begin{equation}
        \Gamma(\hla) = \ll 
	\begin{array}{lcl} 
	{\displaystyle \frac{\gfh \, m_h}{8 \, \sqrt{2} \, \pi} \, 
                \lb \vhlp \rb^2 \, \sqrt{g^{\lep W}} \, {\f_1}^{\lep W}} &
		{\rm for} & (q_h = t, \, X^\pm = W^\pm) \; , \\
	{\displaystyle \frac{\gfh^2 \, m_h}{192 \, \pi^3} \, 
		\lb \vhlp \rb^2 \, {\f_3}^\lep} & 
		{\rm for} & (q_h \not= t, \, X^\pm = \l \bar{\nu}) \; , 
	\end{array}
	\right. 
        \label{eqn:ccd}
\end{equation}
where ${\f_3}^\lep$ accounts the phase space function in 
semi-leptonic decays, i.e.
\begin{equation}
	{\f_3}^x = 1 - 8 \, \mx^2 + 8 \, \mx^6 - \mx^8 
		- 24 \, \mx^4 \, \ln \mx \; , 
	\label{eqn:f3}
\end{equation}
while the QCD corrections to the decays have been omitted.
For $q_h = t$, the QCD corrections are predicted to be tiny.

Now we are discussing the main interest in the paper. As
shown in Eqs. (\ref{eqn:a}) and (\ref{eqn:delta}), non-zero
CP asymmetries arise from non-zero imaginary part of the 
interference terms between the amplitudes and its 
coupling-strengths as well. For top decays in the present model, 
these requirements are satisfied by taking into consideration the 
decay-width in the boson propagator and the complex couplings 
($V, \lp, \lpp$).
This is the reason why one can not neglect the decay-width 
in Eq. (\ref{eqn:f2}) as pointed out before. Calculating $\Delta$ 
defined in Eq. (\ref{eqn:delta}) gives 
\begin{eqnarray}
	\Delta & = & -\frac{
	\gfh^2 \, \fxh^2 \, \mwh^4 \, \mh}{4 \, \pi} \, 
	\sqrt{g^{\l \xi}} \, {\f_1}^{\l \xi} 
	\nonumber \\
	& & \times \sum_{\b_x\not=\b_y}
	\im \[ \( \clibx^\ast \, \chibx \)^\ast \, 
		\( \cliby^\ast \, \chiby \) \]
	\im \[ \( {\f_2}^{\l \xi \b_x} \)^\ast \, 
		{\f_2}^{\l \xi \b_y} \] \; .
	\label{eqn:deltat}
\end{eqnarray}
Using Eqs. (\ref{eqn:a}) and (\ref{eqn:dwhlx}),
CP asymmetry in top decay is found to be
\begin{eqnarray}
	\acp & = & - \sum_{\b_x\not=\b_y} 
	\im \[ \( \clibx^\ast \, \chibx \)^\ast
		\( \cliby^\ast \, \chiby \) \]
	\im \[ \( {\f_2}^{\l \xi \b_x} \) ^\ast
		{\f_2}^{\l \xi \b_y} \] 
	\nonumber \\
	& & \times \ll 
	\sum_\b \lb \cli^\ast \chi \rb^2 
		\lb {\f_2}^{\l \xi \b} \rb^2 
	\right. 
	\nonumber \\
	& & \left. 
	+ \sum_{\b_x\not=\b_y} 
	\re \[ \( \clibx^\ast \chibx \)^\ast 
		\( \cliby^\ast \chiby \) \]
	\re \[ \( {\f_2}^{\l \xi \b_x} \) ^\ast 
		{\f_2}^{\l \xi \b_y} \]	\lr^{-1}.
	\label{eqn:acpt}
\end{eqnarray}

On the other hand, in light-quark decays CP asymmetries 
are largely affected by the hadrons in the initial, 
intermediate and final states \cite{fsi}. Especially, 
as mentioned first, it has been pointed out 
that CP asymmetries of the present class of $B$ decays 
may be enhanced by the long-distance effects of the 
intermediate states with same quark contents as the final 
state, while other intermediate states with different 
quark contents are negligible \cite{bdx}. 
In this case, different amplitude is generated by the 
penguin operator that has different phase with the tree 
one. In our notation, for general hadronic decay $X_{hm} 
\rightarrow X^\prime \rightarrow X_{\l m} \, \xi$ with 
$X^\prime$ is an intermediate state that has the same quark 
content as $X_{\l m} \, \xi$, the amplitude is expressed 
as
\begin{eqnarray}
	\m & = & \sum_\b \ll 
	\( \cli^\ast \, \chi \) \, T_{X_{\l m} \, \xi} +
	\( \clibh^\ast \, \chibh \) \, P_{X_{\l m} \, \xi}
	\right.
	\nonumber \\
	& & \left. + \frac{i}{2} \sum_{X^\prime} \[ 
	\( \cli^\ast \, \chi \) \, T_{X^\prime} +
	\( \clibh^\ast \, \chibh \) \, P_{X^\prime} \] \, 
		\t_{X^\prime} \lr \; ,
	\label{eqn:mre}
\end{eqnarray} 
under an assumption that the rescattering effects can be 
treated perturbatively. $T$ ($P$) denotes the tree (penguin) 
operator, while $\t_{X^\prime}$ denotes the scattering amplitude 
of $X^\prime \rightarrow X_{\l m} \, \xi$. Here, the penguin 
contribution is normalized by the heaviest inner-line 
particle ($q_\hp$) contribution. Hence, $\Delta$ reads 
\begin{eqnarray}
	 \Delta & = & 
	\im \[ \( \cli^\ast \, \chi \)^\ast
		\( \clibh^\ast \, \chibh \) \] 
	\sum_{X^\prime} \im \[ 
	\tilde{T}_{X_{\l m} \, \xi}^\ast \tilde{P}_{X_{\l m} \, \xi} + 
	\tilde{T}_{X^\prime}^\ast \tilde{P}_{X^\prime} 
	\right.
	\nonumber \\
	& & \left. \; \; \; \;  \; \;  + \frac{i}{2} \, 
	\( \tilde{T}_{X_{\l m} \, \xi}^\ast \tilde{P}_{X^\prime}  
	- \tilde{T}_{X^\prime} \tilde{P}_{X_{\l m}\xi}^\ast \) \]  
	\t_{X^\prime} \; ,  
	\label{eqn:deltalq}
\end{eqnarray}
and CP asymmetry in light-quark mode becomes 
\begin{eqnarray}
	\acp & \approx & 
	\sum_{X^\prime} \im \[ \frac{
	\( \tilde{T}_{X_{\l m} \xi}^\ast \tilde{T}_{X^\prime}^\ast \) \,  
		\t_{X^\prime}}{
		\lb \tilde{T}_{X_{\l m} \xi} \rb^2 + 
		\lb \bar{\tilde{T}}_{X_{\l m} \xi} \rb^2 } 
	\right.
	\nonumber \\
	& & \left. \; \; \; \; \; \; \; \; \; \; \; \; \times
	\( \frac{\tilde{P}_{X_{\l m} \xi}}{\tilde{T}_{X^\prime}^\ast} + 
	\frac{\tilde{P}_{X^\prime}}{\tilde{T}_{X_{\l m} \xi}^\ast} + 
	\frac{i}{2} \( 
	\frac{\tilde{P}_{X^\prime}}{\tilde{T}_{X^\prime}^\ast} - 
 	\frac{\tilde{P}_{X_{\l m} \, \xi}}{\tilde{T}_{X_{\l m} \, \xi}^\ast}\) 
	\) \] 
	\nonumber \\
	& & \; \; \; \; \; \; \times \ll 
	\begin{array}{ll}
	{\displaystyle 
		\im \[ \frac{{V_{\l\hp}}^\ast \, V_{h\hp}}{
			{V_{\l j}}^\ast V_{hj}} \]} 
		& {\rm for \; type \; 1} \; , \\
	{\displaystyle 
		\im \[ \frac{{V_{\l\hp}}^\ast \, V_{h\hp}}{
			\clifp^\ast \, \chifp} \] }
		& {\rm for \; type \; 2} \; ,
	\end{array}
	\right. 
	\label{eqn:acplq}
\end{eqnarray}
if the total decay-width is approximately dominated by tree 
operator. Here $\sf^\prime$ denotes the lightest sfermion 
and $\bar{\tilde{T}}$ is a complex conjugate of $\tilde{T}$.
A tilde means 
\begin{equation}
	\tilde{T} \equiv \ll 
	\begin{array}{ll}
	{\displaystyle 
	T^W \, \( 1 + \sum_\sf 
	\frac{\clif^\ast \, \chif}{{V_{\l j}}^\ast V_{hj}} 
	\frac{{\f_2}^{\l \xi \sf}}{{\f_2}^{\l \xi W}} \) }
	& {\rm for \; type \; 1} \; , \\
	{\displaystyle 
	T^{\sf^\prime} \, \( 1 + \sum_{\sf\not=\sf^\prime} 
	\frac{\clif^\ast \, \chif}{\clifp^\ast \, \chifp} 
	\frac{{\f_2}^{\l \xi \sf}}{{\f_2}^{\l \xi \sf^\prime}} \) }
	& {\rm for \; type \; 2} \; ,
	\end{array}
	\right.
	\label{eqn:tbar}
\end{equation}
while 
\begin{equation}
	\tilde{P} \equiv P^W \, \( 1 + \sum_\sf 
	\frac{\clif^\ast \, \chif}{{V_{\l\hp}}^\ast V_{h\hp}} 
	\frac{P^\sf}{P^W} \) \; ,
	\label{eqn:pbar}
\end{equation}
for both types. $T^\b$ ($P^\b$) denotes the $\b-$boson mediated 
tree (penguin) operator. Since the matrix elements 
depend on the hadronic states, 
$\<\tilde{T}_{X_{\l m}\xi}\>$ $\(\<\tilde{P}_{X_{\l m} \xi}\>\)$ 
is in general different with 
$\<\tilde{T}_{X^\prime}\>$ $\(\<\tilde{P}_{X^\prime}\>\)$. 

Remark that the results in Eqs. (\ref{eqn:dwhlx}), 
(\ref{eqn:deltat}) and (\ref{eqn:deltalq}), at least numerically, 
are not altered so much by the diagonalization of squarks 
($\sq_L, \sq_R$),
although in fact, squarks are essentially mixed each other due to 
large Yukawa coupling of their partner quarks in the MSSM \cite{diagonal}. 
So for a rough order estimation and also reducing the model
dependence on the diagonalization, it is better to use the 
weak eigenstate as it is. Imagine the process is through a
squark mediated diagram, then we can appreciate this point in 
two extreme cases. First case is when the masses are almost 
decoupled, then the branching-ratio will be quadruple, while
the CP asymmetry will be reduced by half. On the contrary, 
when the mass difference is extremely large, one can neglect
the large ones because the contribution will be supressed 
by inverse of its mass square. 

Now we are ready to make numerical analysis for the 
branching-ratios and CP asymmetries. Many authors have 
extracted some direct and indirect bounds for the coupling 
strengths in Eqs. (\ref{eqn:lglp}) and (\ref{eqn:lglpp}), 
i.e. $\lp$ and $\lpp$. The bounds can be seen in Tab. 1 of ref. 
\cite{constraint}. However, until now there is still no 
rigid constraints for $\lpp$. Moreover, since one of the 
$B-$ and $L-$parity is still possible solutions to maintain 
a stable proton and allow for $\rp-$violation as well, we assume that 
only one of these symmetries has been violated. Next, we 
consider only the must-be lightest sfermion for each sector 
and neglect the other heavier sfermions because its 
contributions should be suppressed. Hence, 
the analysis is simplified and can be done 
in a general way, i.e. it is sufficient to consider $W-$ and 
one $\sf-$mediated diagrams for type 1, or only single  
$\sf-$mediated diagram for type 2.

For the branching-ratios of charged-current decays, 
we put $\br(b \rightarrow c \, \l \bar{\nu}) = 0.103$ 
and $\br(\tbw) \sim 1$ by assuming the mode to be dominant 
in top quark decays. Also use the experimental results, 
$m_u = 6$(MeV), $m_c = 1.3$(GeV), $m_t = 180$(GeV), 
$m_d = 10$(MeV), $m_s = 200$(MeV), $m_b = 4.3$(GeV), 
$m_W = 80.33$(GeV), $\Gamma_W = 2.07$(GeV) and 
the Wolfenstein parameters of CKM matrix 
$(A,\lambda,\rho,\eta) = (0.86,0.22,0.3,0.34)$ \cite{pdg}, 
In figure captions, the couplings are redefined as 
$C_W \equiv V_{\l j}^\ast V_{hj}$ and 
$C_\sf \equiv {\lambda_{ijk}^\prime}^\ast 
\lambda_{i^\prime j^\prime k^\prime}^\prime$ or 
${\lambda_{ijk}^{\prime\prime}}^\ast 
\lambda_{i^\prime j^\prime k^\prime}^{\prime\prime}$ respectively.
The size of $C_\sf$ in some figures is fixed to be $\sim0.015$
that is reliable enough for most coupling-strengths listed in 
\cite{constraint}.
In all figures, for the must-be lightest sfermion mass we put 
$\frac{1}{2} m_t < m_\sf < m_t$ that satisfies the kinematical 
requirement mentioned below Eq. (\ref{eqn:f2}). This region is 
still above the lower bound from the LEP experiments. Further,
we put $\Gamma_\sf \sim 2 \Gamma_W$ for whole region of 
sfermion masses. For the narrow region of masses under consideration, 
this approximation is good although in general the decay-width 
must be dependent on the mass.
The phase of complex coupling $C_\sf$ is defined as
\begin{equation}
	C_\sf \equiv \lb C_\sf \rb \, e^{i \theta} \; .
	\label{eqn:theta}
\end{equation}

\begin{figure}[t]
\begin{center}
\begin{tabular}{cc}
       	\leavevmode\psfig{file=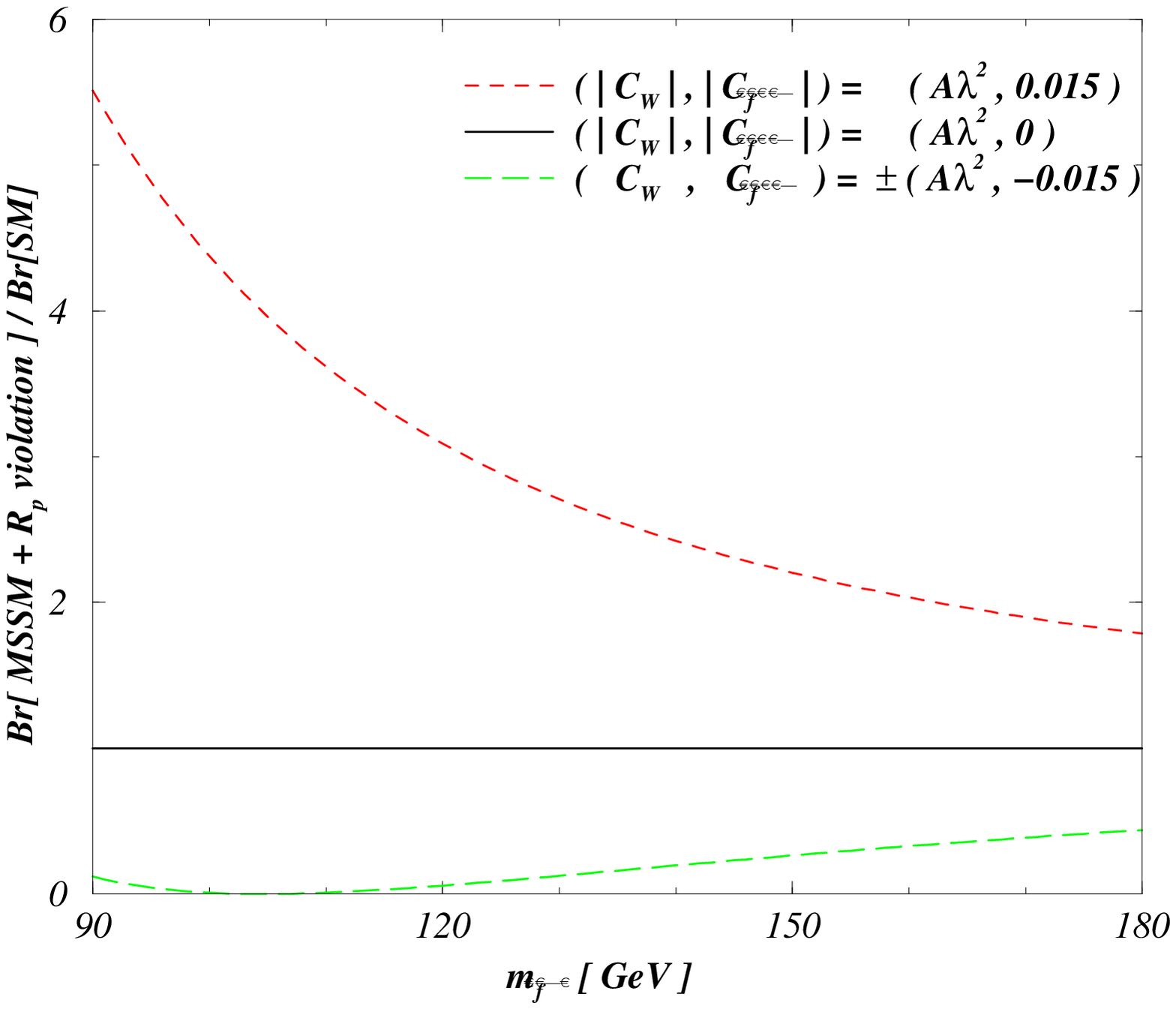,height=8cm,width=7cm}
	&
       	\leavevmode\psfig{file=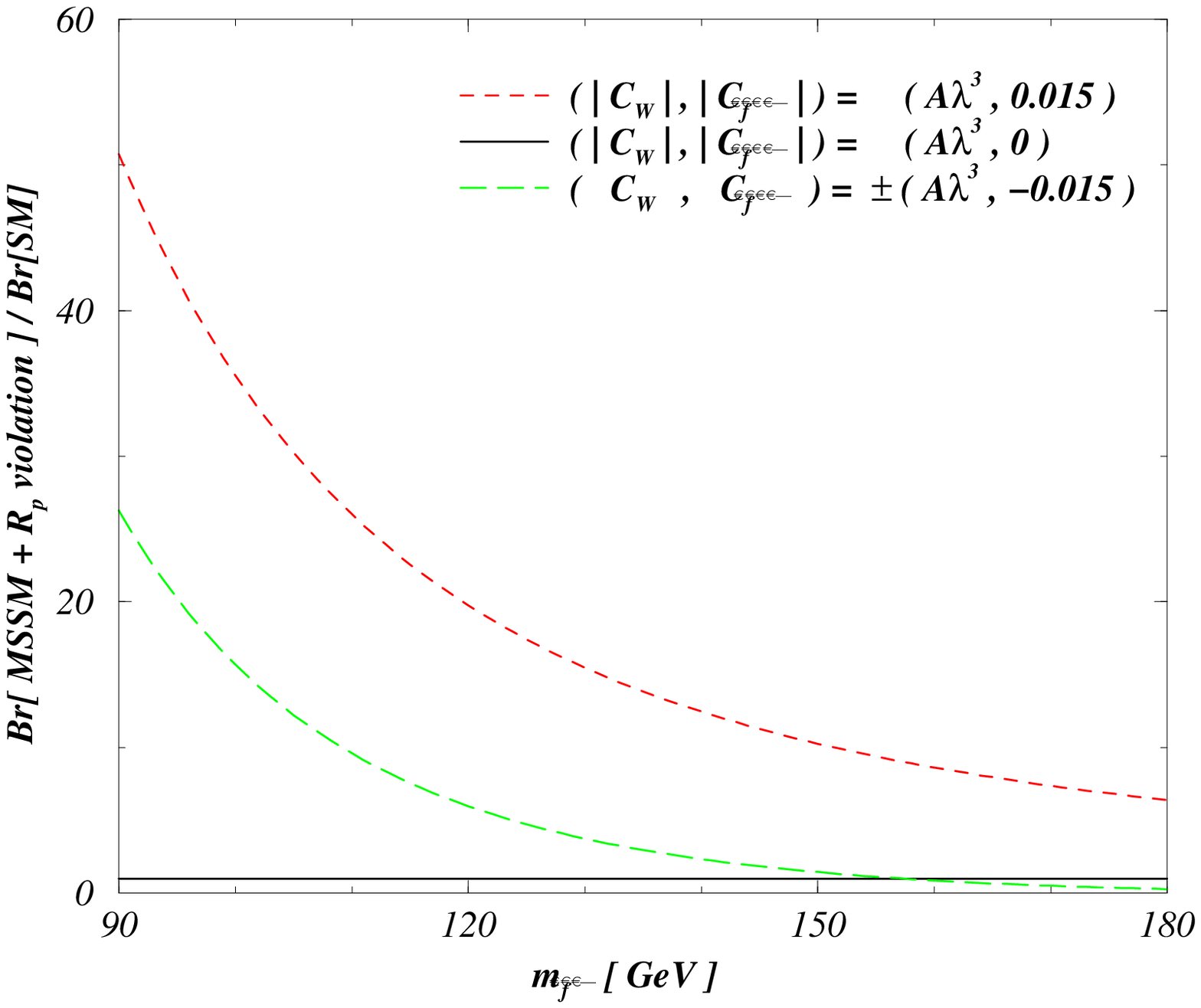,height=8cm,width=7cm}
	\end{tabular}
	\caption{Ratio of the branching-ratio of 
	$b \rightarrow s \, J/\psi$ (left), and 
	$b \rightarrow d \, J/\psi$ (right).}
	\label{fig:rbrlq}
	\end{center}
\end{figure}

\begin{figure}[h]
\begin{center}
\begin{tabular}{cc}
       	\leavevmode\psfig{file=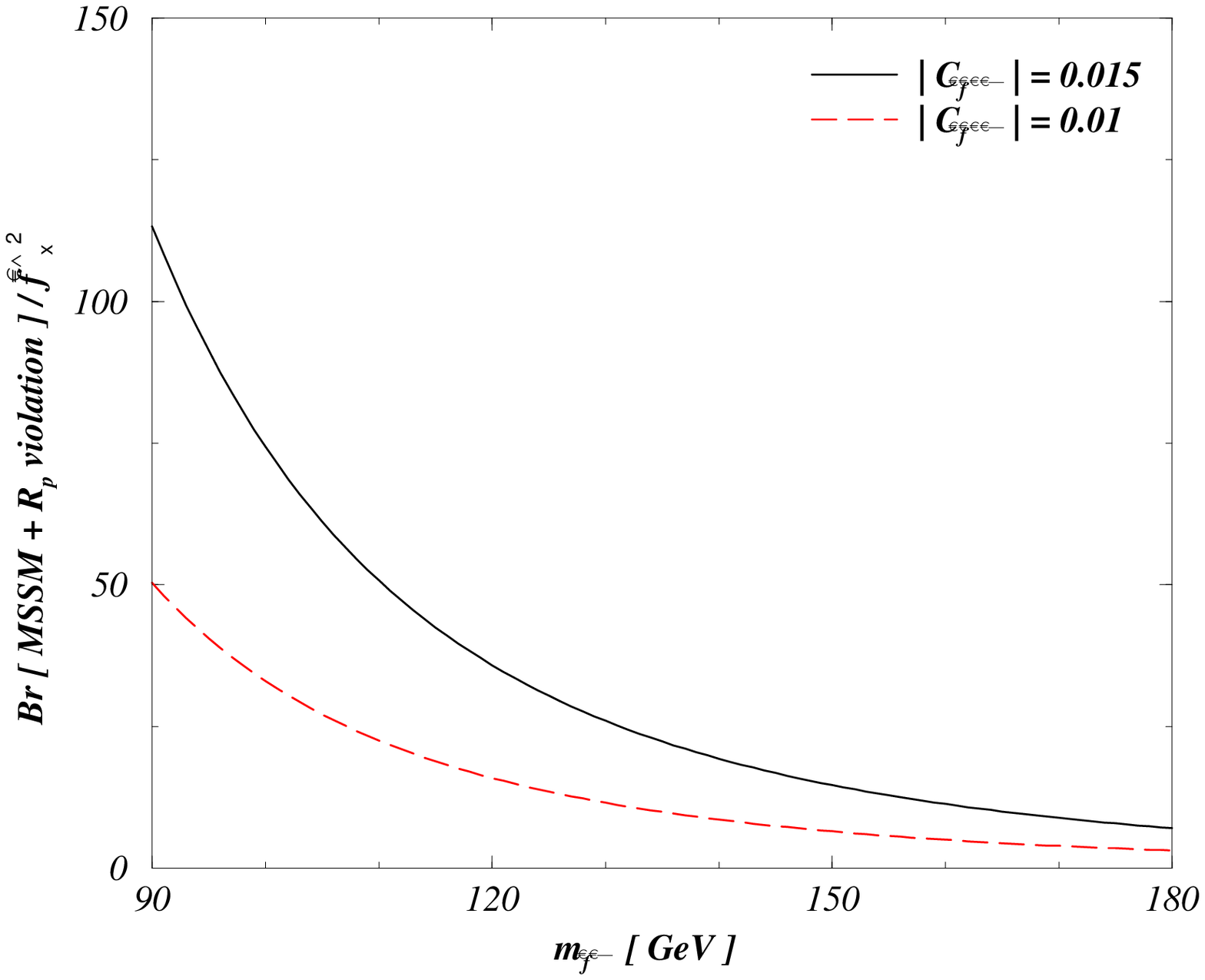,height=8cm,width=7cm}
	&
       	\leavevmode\psfig{file=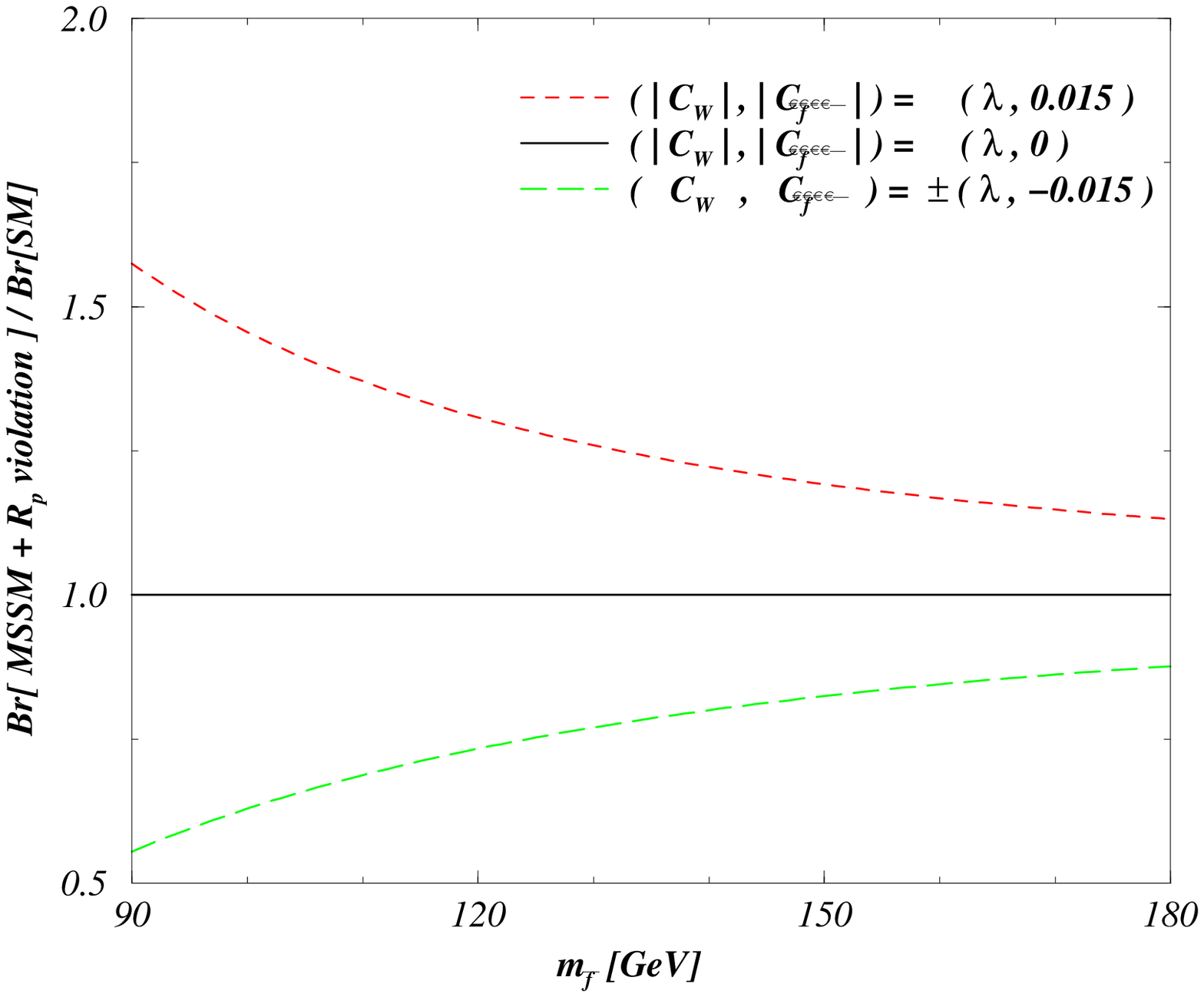,height=8cm,width=7cm}
	\end{tabular}
	\caption{The branching-ratio of 
	$b \rightarrow d(s) \, \phi$ or 
	$b \rightarrow d(s) \, \rho(\omega)$ (left), 
	and ratio of the branching-ratio of 
	$s \rightarrow d \, \rho(\omega)$ or 
	$c \rightarrow u \, \phi$ (right).}
	\label{fig:brlq}
	\end{center}
\end{figure}

Since large branching-ratios of light-quark decays in the SM 
are favored, it is better to describe the ratio of SM and 
MSSM with $\rp-$violation cases. Then the unknown parameter $\fxh$
will be eliminated as shown in Figs. \ref{fig:rbrlq} and 
\ref{fig:brlq} (the right one). However, in case of either type 
1 modes, or type 2 modes with tiny $C_W$ ($C_\sf \gg C_W \sim 0$),  
one must plot the branching-ratios itself as depicted in the left 
figure in Fig. \ref{fig:brlq} with leaving $\fxh$ as unknown. 
Note that in Fig. \ref{fig:brlq}, it seems significant differences 
between neither $b \rightarrow d(s) \, \phi$ with 
$b \rightarrow d(s) \, \rho(\omega)$ nor   
$s \rightarrow d \, \rho(\omega)$ with $c \rightarrow u \, \phi$.
On the other hand, for CP asymmetries in light-quark decays, 
a rough prediction can be performed simply by using Eqs. 
(\ref{eqn:f2}) and (\ref{eqn:tbar}). In general CP asymmetry for 
type 1 will be changed by a factor of 
\begin{equation}
	\acp \approx {\acp}^{\rm SM} \, \( 1 + 
		\frac{a \, C_\sf}{C_W} 
		\frac{{m_W}^2}{{m_\sf}^2} \)^{-1} \; ,
	\label{eqn:acpmssm}
\end{equation}
since $\<\tilde{P}\> \approx \<P^W\>$ for $C_\sf$ around 
the present value \cite{2hdm,mssmrp}. For example, let us 
consider CP asymmetry in the decay $b \rightarrow d \, J/\psi$. 
In the present model it will be changed to be 
$\acp \approx (-14 \sim 2) \times {\acp}^{\rm SM}$ for 
$C_\sf = 0.015$ and various set of $(\theta,m_\sf)$.
On the other hand, in the framework of SM with including 
the long-distance effects, e.g. 
$B^- \rightarrow D^0 \, D^- \rightarrow \pi^- \, J/\psi$, 
the value has been predicted to be ${\acp}^{\rm SM} \sim 1\%$ \cite{bdx}. 
The prediction for other light-quark modes can be accomplished 
by the same procedure respectively. Remark that the procedure here 
is not requiring any kinematical condition like before, i.e.
$m_\b < m_h$.

\begin{figure}[t]
       	\begin{center}
       	\leavevmode\psfig{file=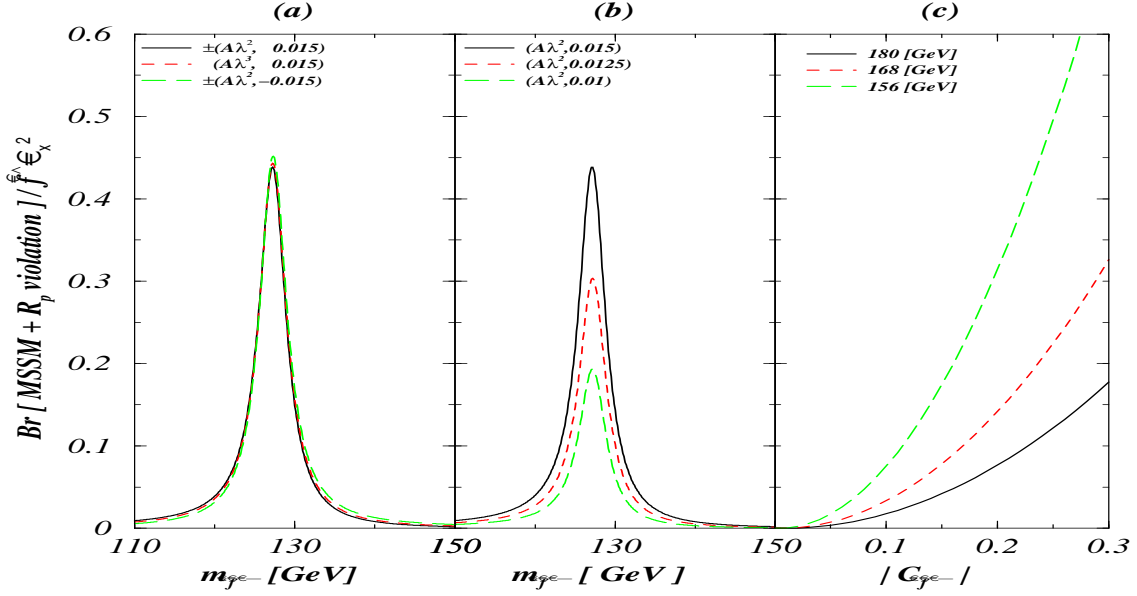,height=8cm,width=15cm}
	\caption{The branching-ratios of 
	$t \rightarrow u \, \phi(\Upsilon)$ and 
	$t \rightarrow c \, \phi(\Upsilon)$ 
	for various 
	(a) $(C_W, C_\sf)$, 
	(b) $C_\sf$ with $C_W=A\lambda^2$, and  
	(c) $m_\sf$ with $C_W=A\lambda^2$.}
	\label{fig:brt}
       	\end{center}
\end{figure}

\begin{figure}[h]
       	\begin{center}
       	\leavevmode\psfig{file=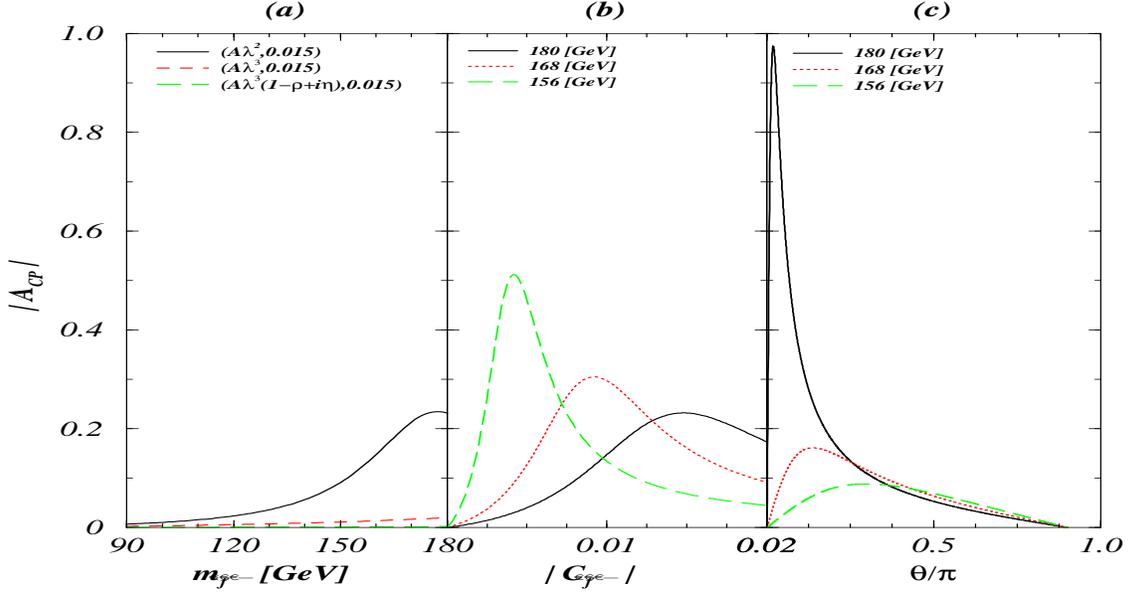,height=8cm,width=15cm}
	\caption{The CP asymmetries in  
	$t \rightarrow u \, \phi(\Upsilon)$ and 
	$t \rightarrow c \, \phi(\Upsilon)$ for various 
	(a) $(C_W, C_\sf)$ with $\theta=\pi/6$, 
	(b) $m_\sf$ with $\theta=\pi/6$, and 
	(c) $\theta$ with $(|C_W|, |C_\sf|)=(A\lambda^2,0.015)$.}
	\label{fig:acpt}
       	\end{center}
\end{figure}

In top decays, the dependences on $m_\l$ and $m_\xi$ are drastically
suppressed. It makes the discrepancies between different modes are 
almost coming from CKM matrix elements. So one can describe them 
generally as depicted in Figs. \ref{fig:brt} and \ref{fig:acpt}.
From the figures, the sfermion contributions will be maximum 
near the resonances. An interesting behaviour 
appears in the figure (b) in Fig. \ref{fig:acpt}, that is
in the small coupling region light sfermions are favored 
to obtain large CP asymmetry and vice versa.

In conclusion, the class of decay modes $\hlx$ have been 
studied in the framework of the MSSM without $\rp$. 
The study has been done for top and light-quark decays 
simultaneously, and focused on the CP asymmetries for 
$\frac{1}{2} m_t < m_\sf < m_t$. It is 
shown that, CP asymmetries in top decays can be induced 
by taking into account the decay-widths of the 
exchanged-bosons, while in light-quark decays it can be generated 
due to the long-distance effects as usual. The sfermion 
contributions due to the new interactions in $\rp-$violation 
superpotential change the branching-ratios and CP asymmetries 
significantly. Both measurements are very sensitive to the 
coupling-strengths and sfermion masses as well, that makes 
the modes to be good probes to search for the $\rp-$violation 
in the MSSM. Finally, although the decays are suffered from the 
small Yukawa couplings compared with some supersymmetric 
productions, they have double kinematic reach that makes them to 
be better for achieving precise measurements. Therefore, combining 
the various production and decay modes will lead to a wide range 
of potential signals to search for the $\rp-$violation in the 
MSSM.

LTH would like to thank Ministry of Education and Culture of
Japan for the financial support under Monbusho Fellowship 
Program.

\clearpage

\clearpage

\end{document}